\documentclass[final,3p,times,twocolumn]{elsarticle}
\usepackage{amsmath,amssymb,graphicx}
\usepackage{color}
\usepackage{multirow}
\usepackage{float}
\usepackage{subfigure}
\usepackage{rotating}
\usepackage{wasysym}

\pdfoutput=1

\newcommand{\nc}{\newcommand}
\nc{\postscript}[2]
{\setlength{\epsfxsize}{#2\hsize}\centerline{\epsfbox{#1}}}
\nc{\non}{\nonumber}
\nc{\hc}{\hbox {h.c.}} \nc{\re}{\hbox {Re}} 
\nc{\mev}{\hbox {MeV}} \nc{\gev}{\;\hbox {GeV}} \nc{\tev}{\;\hbox {TeV}}
\def\lsim{\mathrel{\raise.3ex\hbox{$<$\kern-.75em\lower1ex\hbox{$\sim$}}}}
\def\gsim{\mathrel{\raise.3ex\hbox{$>$\kern-.75em\lower1ex\hbox{$\sim$}}}}

\nc{\etal}{{\it et al.}}
\nc{\Lsp}{\;\;\;\;\;\;\;\;\;\;}  \nc{\LLLsp}{\lspace \lspace}
\nc{\lsp}{\;\;\;\;\;\;}
\nc{\spac}{\;\;\;}
\nc{\noi}{\noindent}
\nc{\beq}{\begin{equation}}   \nc{\eeq}{\end{equation}}
\nc{\bea}{\begin{eqnarray}}   \nc{\eea}{\end{eqnarray}}
\nc{\baa}{\begin{array}}      \nc{\eaa}{\end{array}}
\nc{\bit}{\begin{itemize}}    \nc{\eit}{\end{itemize}}
\nc{\ben}{\begin{enumerate}}  \nc{\een}{\end{enumerate}}
\nc{\bce}{\begin{center}}     \nc{\ece}{\end{center}}

\def\sq2{\sqrt{2}}

\def\ph{\varphi}

\def\m4{m^4(\ph)}
\def\mn2{m_n^2}

\def\v5{V^{(5)}}

\def\baa{\begin{array}}
\def\eaa{\end{array}}

\journal{Physics Letters B}

\begin{document}

\begin{frontmatter}


\title {Effects of a Real Singlet Scalar on Veltman Condition}

\vskip 20pt

\author[label1]{Canan Nurhan Karahan}
\ead{cananduzturk@iyte.edu.tr}
\author[label1]{Beste Korutlu}
\ead{bestekorutlu@iyte.edu.tr}
\address[label1]{Department of Physics, \.{I}zmir Institute of Technology, IZTECH, TR35430, \.{I}zmir, Turkey}

\date{\today}

\begin{abstract}
We revisit the fine-tuning problem in the Standard Model (SM) and show the modification in Veltman condition by virtue of a minimally-extended particle spectrum with one real SM gauge singlet scalar field. We demand the new scalar to interact with the SM fields through Higgs portal only, and that the new singlet to acquire a vacuum expectation value, resulting in a mixing with the CP-even neutral component of the Higgs doublet in the SM. The experimental bounds on the mixing angle are determined by the observed best-fit signal strength $\sigma/\sigma_{\rm SM}$. While, the one-loop radiative corrections to the Higgs mass squared, computed with an ultraviolet cut-off scale $\Lambda$, comes with a negative coefficient, the quantum corrections to the singlet mass squared acquires both positive and negative values depending on the parameter space chosen, which if positive might be eliminated by introducing singlet or doublet vector-like fermions. However, based upon the fact that there is mixing between the scalars, when transformed into the physical states, the tree-level coupling of the Higgs field to the vector-like fermions worsen the Higgs mass hierarchy problem. Therefore, the common attempt to introduce vector-like fermions to cancel the divergences in the new scalar mass, might not be a solution, if there is mixing between the scalars.
 \end{abstract}

\end{frontmatter}

\section{Introduction}
\label{sec:intro}
The discovery of the fundamental particle of mass $m_h=125.9\pm 0.4$ GeV at the Large Hadron Collider (LHC) \cite{Aad:2012tfa, Chatrchyan:2012ufa}, highly likely to be the Higgs boson of the Standard Model (SM), settles the experimental validation for the long sought missing piece of the SM. The absence of any new physics signals by the first 20 fb$^{-1}$ of data from LHC operating at 8 TeV, on the other hand, casts doubts on the relevance of our notion of the naturalness problem. At this very moment of the shut down of LHC, after a glimpse of what may be the Higgs boson, understanding the hierarchy problem seems crucial. In an effective field theory approach with an ultraviolet cut-off $\Lambda$, the Higgs self energy receives quadratically divergent corrections from loop diagrams such that
\bea
m_h^2=(m_h^2)_{\rm bare}+\mathcal{O}(\lambda_H, g_i^2)\Lambda^2,
\eea
where $m_h=\sqrt{2\lambda_H}\upsilon_H$ is the physical Higgs mass, $\lambda_H$ is the Higgs self coupling, and $g_i$ are the renormalized couplings of the SM. Hence, the natural scale for the Higgs mass is $\mathcal{O}(\Lambda)$, and is unnatural for it to be less than the ultraviolet cut-off of the theory, which could be as high as the Planck Scale ($M_{\rm Pl}\sim 10^{19}$ GeV).

Originally studied by Veltman \cite{Veltman:1980mj}, the SM one-loop condition of the quadratic divergences reads
\bea\label{eq:veltman}
\delta m_h^2=\frac{\Lambda^2}{16 \pi^2}\left(6\lambda_H+\frac{9}{4}g^2+\frac{3}{4}g'^2-6g_t^2\right),
\eea
where $g$ and $g'$ are the $SU(2)_L$ and $U(1)_Y$ gauge couplings of the SM, respectively, and $g_t=\sqrt{2}m_t/\upsilon_H$ ($\upsilon_H=246$ GeV is the vacuum expectation value (VEV) of the Higgs field) is the top quark Yukawa coupling. Since the contributions to the Veltman condition (VC) by other fermions are considerably small compared to the one by top quark, they are not taken into account. The VC demands that, the quadratically divergent terms above adds up either to zero, or to a very small value by virtue of some symmetry of the model. Now that we know all the masses, based on the requirement that $|\delta m_h^2|/m_h^2\leq 1$, the VC is in conflict with the experimental data for $\Lambda>780$ GeV. There have been various attempts to protect Higgs mass from destabilizing by introducing a new set of particles and interactions. Chief among them is Supersymmetry, which solves the gauge hierarchy problem introducing supersymmetric partners of the SM particles with masses around TeV (see \cite{Martin:1997ns} for a review on supersymmetry and \cite{Masina:2013wja} as a recent work on VC in a High-Scale Supersymmetric model). So far, however, no compelling sign of experimental evidence for supersymmetry has been found, in the searches at the Large Hadron Collider (LHC) \cite{Feng:2009te}. Other possible solutions are Little Higgs Models \cite{ArkaniHamed:2002qy}, Large and/or Warped Extra Dimensions \cite{ArkaniHamed:1998rs}. There exists also a very different perspective, motivated by anthropic considerations \citep{Agrawal:1997gf}, arguing that the notion of the naturalness of the weak scale should be abandoned.

In the present work, we show the effect of extending the scalar sector of the SM minimally, by introducing a real gauge singlet scalar field that only couples with the Higgs doublet, on the VC. The phenomenology of singlet scalars has been extensively studied by \cite{Bjorken:1991xr} as a hidden sector, by \cite{Barger:2007im,Barger:2008jx} as dark matter candidate, and their effect on the stability condition has been previously discussed in \cite{Kundu:1994bs,Chakraborty:2012rb}. We discuss also the radiative corrections for real singlet scalar, and as opposed to what has been found in the literature, we show that even a small mixing angle might result interesting results, and that introducing singlet or doublet vector fermions might not be a valid scenario for both scalars in the model (Higgs boson and singlet scalar) in the presence of mixing between the two.

\section{The Model}
We consider the simplest extension of the SM by introducing a real singlet scalar field $S$. We impose an additional ${\mathbb{Z}}_2$ symmetry under which $S$ is odd. The most general, renormalizable, symmetric Lagrangian density reads,
\bea
\mathcal{L}_{HS}&=&(D_{\mu}H)^{\dagger}D^{\mu}H+\frac{1}{2}\partial_{\mu}S\partial^{\mu}S-V_{H\!S},
\eea
where
\bea\label{eq:pot}
V_{\!H\!S}\!\!=\!\!\mu_H^2\!H^{\dagger}\!H\!+\!\lambda_H(\!H^{\dagger}\!H\!)^2\!\!+\!\frac{\mu_S^2}{2}\!S^2\!+\!\!\frac{\lambda_S}{4}\!S^4\!+\!\frac{\lambda}{2}\!H^{\dagger}\!H\!S^2,
\eea
is the potential, and $H$ is the SM Higgs doublet. The potential is bounded from below for $\lambda_H>0$ and $\lambda_S>0$, and the minimum of the potential breaks the electroweak symmetry spontaneously via non-zero vacuum expectation value (VEV) of the Higgs doublet, $\langle H \rangle=\upsilon_H/\sqrt{2}$, generating masses for the SM particles. Depending on whether the real singlet scalar develops a VEV or not, the VC takes different forms. In the subsequent analysis, we focus on the former. However, in explaining the details of our model, we will follow a pedagogical strategy, and include also the details of the model where $S$ is not developing a VEV.\\

\subsection{Singlet VEV Vanishes}
If the real singlet scalar does not acquire a VEV, minimum of the potential occurs at
\bea
\upsilon_H^2= -\frac{\mu_H^2}{\lambda_H},\qquad \upsilon_S^2=0.
\eea
Using the parametrization of the Higgs field above the vacuum as
\bea\label{eq:Higgs}
H=\frac{1}{\sqrt{2}}\left(\baa{c}
H_{3}+i H_{4}\\
\upsilon_H+H_{1}+i H_{2}
\eaa
\right),
\eea
the mass squared values of $H_1\equiv h$ (the CP-even neutral component of the Higgs doublet) and the $S$ fields (for later convenience, we use $\sigma$ for the corresponding physical real singlet scalar) are obtained as
\bea\label{eq:nomix}
m_{h}^2=2\lambda_H\upsilon_H^2, \qquad m_{\sigma}^2=\mu_S^2+\frac{\lambda}{2}\upsilon_H^2.
\eea
In this scenario, with $\upsilon_S=0$, there is no mixing between the Higgs and the new scalar field. Therefore, they appear naturally in their mass eigenstates. Moreover, as it can be seen in Eq. (\ref{eq:pot}), despite there is no mixing, the quartic interaction term $\frac{\lambda}{4}\sigma^2h^2$ between the $\sigma$ and $h$ fields still exist as long as $\lambda\neq 0$. \\

\subsection{Singlet VEV does not vanish}
If the real singlet scalar acquires a non-zero VEV, the minimum of the potential occurs at
\bea
\upsilon_H^2= \frac{4\lambda_S \mu_H^2-2\lambda \mu_S^2}{\lambda^2-4\lambda_{H}\lambda_S},\!\!\qquad\!\! \upsilon_S^2=\frac{4\lambda_H \mu_S^2-2\lambda \mu_H^2}{\lambda^2-4\lambda_{H}\lambda_S}.
\eea
Using the parametrization of the Higgs field above the vacuum as in Eq. (\ref{eq:Higgs}), and for the real singlet scalar as $S=\upsilon_S+S$
we obtain the mass squared mixing matrix for the fields $H_1$ and $S$
\bea
M^2_{H_1,S}=\left(\baa{cc}
2\lambda_H\upsilon_H^2 & \frac{\lambda}{2}\upsilon_H\upsilon_S \\
\frac{\lambda}{2}\upsilon_H\upsilon_S & 2\lambda_S\upsilon_S^2
\eaa
\right).
\eea
After diagonalization, the mass matrix yields the masses of the physical scalar $h$ (Higgs field) and $\sigma$ fields as follows
\bea\label{eq:mh2msig2}
m_h^2\!\!=\!\upsilon_H^2\!\left(\!\lambda_H\!+
\!\lambda_S\upsilon_{SH}^2\!\!
+\!\!\sqrt{\!(\lambda_H\!-\!\lambda_S\upsilon_{SH}^2)^2
\!\!+\!\!\frac{\lambda^2}{4}\!\upsilon_{SH}^2}\right)\!,&&\non\\
m_{\sigma}^2\!\!=\!\upsilon_H^2\!\left(\!\lambda_H\!+
\!\lambda_S\upsilon_{SH}^2\!\!
-\!\!\sqrt{\!(\lambda_H\!-\!\lambda_S\upsilon_{SH}^2)^2
\!\!+\!\!\frac{\lambda^2}{4}\!\upsilon_{SH}^2}\right)\!,&&
\eea
where $\upsilon_{SH}=\frac{\upsilon_S}{\upsilon_H}$.
The mixing angle in terms of model parameters reads
\bea\label{eq:tantheta}
\tan 2\theta=\frac{\lambda \,\upsilon_{SH}}{\lambda_S\upsilon_{SH}^2-\lambda_H}.
\eea
As expected, the mixing angle $\theta$ is proportional to $\lambda$.

Below, we study in detail the latter case, to analyse the contributions to the VC coming from the singlet sector.
%
\section{Phenomenology}
The presence of one real singlet scalar field $S$, with VEV $\upsilon_S$, modifies the VC for the Higgs mass, not only due to its direct coupling to SM Higgs doublet, but also its mixing with the neutral, CP-even component of the doublet, which allows tree-level interactions of the new scalar with the SM fields. We carry our calculations up to one-loop order.

The VC is modified with the addition of one real singlet scalar field as follows (See Appendix for the vertex factors):
\bea\label{eq:delmh2}
\delta m_h^2\!\!\!&=&\!\!\frac{\Lambda^2}{16 \pi^2}\bigg[\!\cos^4\!\theta\left(\!\frac{\lambda}{2}\!\!+3\lambda_H\!\right)\!\!+\!\sin^4\!\theta\left(\!\frac{\lambda}{2}+\!3\lambda_S\!\right)\!\non\\
&+&\!\sin^2\!2\theta\left(\frac{3\lambda_H}{4}\!\!+\!\!\frac{3\lambda_S}{4}\!\!+\!\!\frac{\lambda}{4}\!\right)+\sin^2\!\theta\left(\frac{3\lambda}{2}\right)\non\\
&+&\cos^2\theta\!\left(3\!\lambda_H\!\!+\!\frac{9}{4}g^2\!\!+\!\frac{3}{4}g'^2\!\!-\!6g_t^2\!\right)\bigg].
\eea
%
%
%
In the limit of no mixing (i.e., $\cos\theta\rightarrow 1$, $\sin\theta\rightarrow 0$) the modified VC reduces to the original one given in Eq. (\ref{eq:veltman}), except the term $\frac{\Lambda^2}{16\pi^2}\frac{\lambda}{2}$, which appears regardless of the mixing, and disappears if and only if $\lambda=0$. The reason is that vanishing of mass mixing ($\theta\rightarrow 0$) does not guarantee vanishing of the quartic mixing ($\frac{\lambda}{2}H^{\dagger}HS^2$). The latter gives contribution to VC even when $\theta\rightarrow 0$.

\begin{figure}[h]
\includegraphics[scale=0.69]{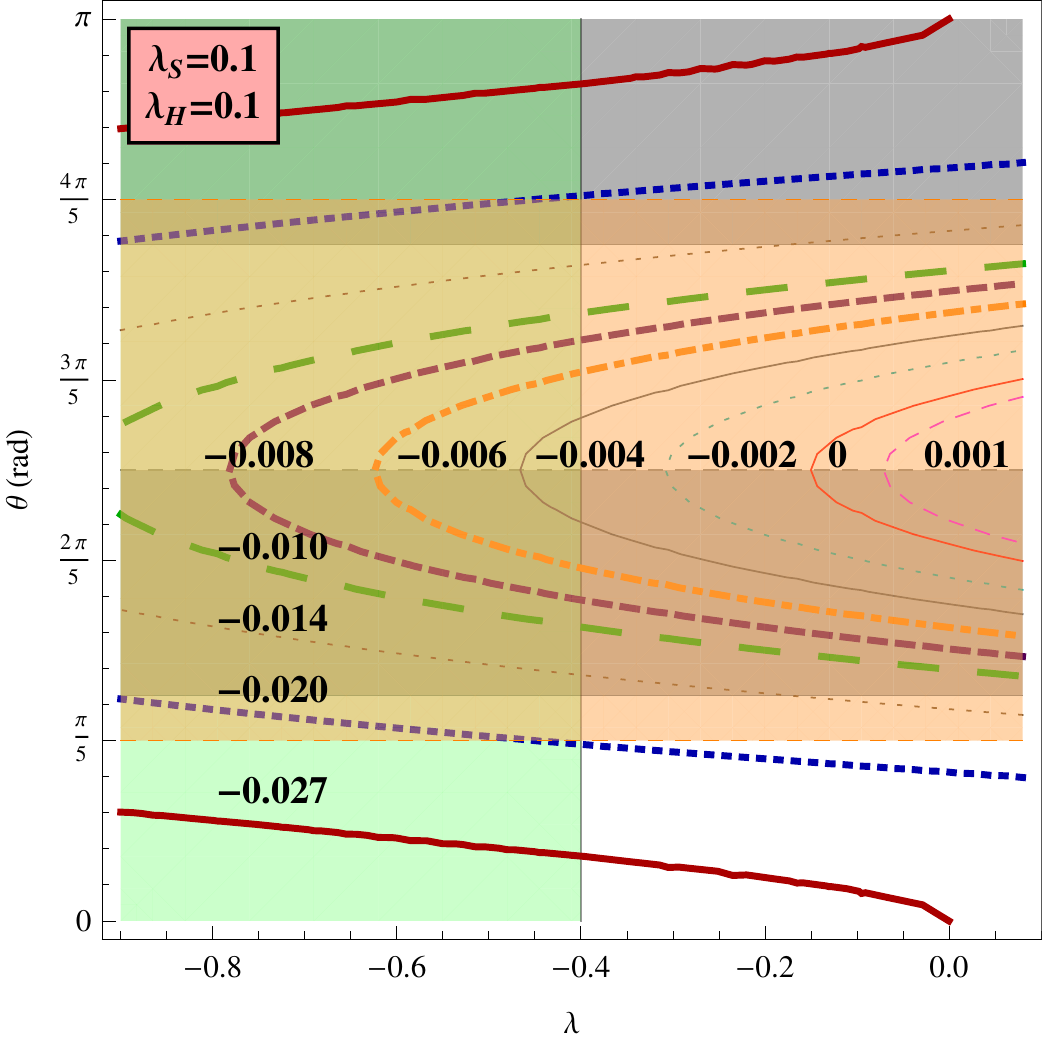}
\caption{VC for the Higgs boson for $\lambda_H=\lambda_S=0.1$. Contours represent $\delta m_h^2$ in units of $\Lambda^2$. The shaded regions are excluded either because they represent non-physical situations or by the experiments. The light-gray shaded region is where $\upsilon_{SH}=\upsilon_{S}/\upsilon_H< 0$, light-green shaded region is where $m_{\sigma}^2$ is negative, and light-orange shaded region is ruled out at the LHC via the best-fit signal strength measurements $\sigma/\sigma_{SM}=\cos^2\theta=0.87\pm 0.23$. In this plot, there is always a region satisfying $123\, {\rm GeV}\leq m_h \leq 129\, {\rm GeV}$ and the region for singlet mass is $ m_{\sigma} \leq 108 \,{\rm GeV}$. Introducing vector fermions entails the undesired effect of making the negative $\delta m_h^2$ even more negative.}
\label{fig:Veltman1}
\end{figure}
\begin{figure}[h]
\includegraphics[scale=0.69]{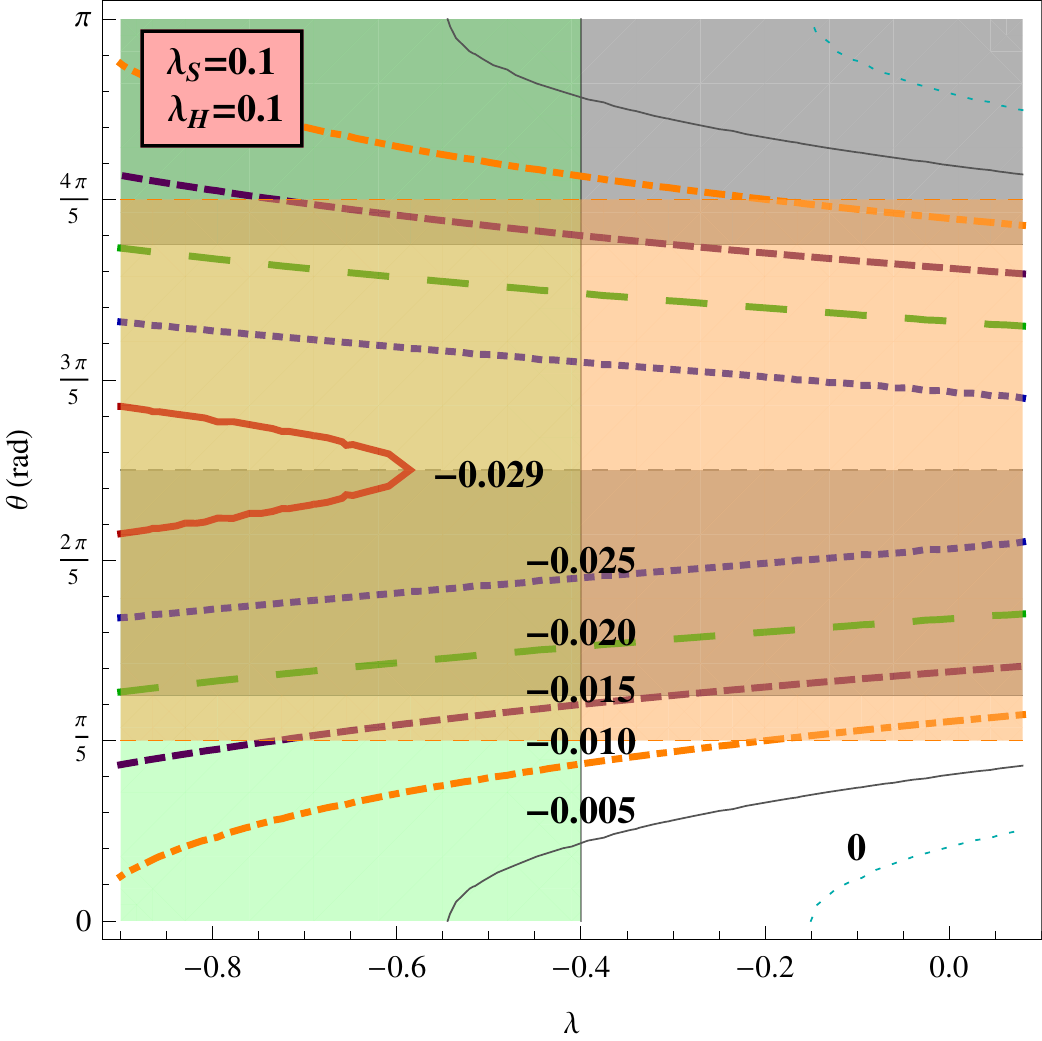}
\caption{VC for the real singlet scalar for $\lambda_H=\lambda_S=0.1$. Contours represent $\delta m_{\sigma}^2$ in units of $\Lambda^2$. The shaded regions are excluded either because they represent non-physical situations or by the experiments. The light-gray shaded region is where $\upsilon_{SH}=\upsilon_{S}/\upsilon_H< 0$, light-green shaded region is where $m_{\sigma}^2$ is negative, and light-orange shaded region is ruled out at the LHC via the best-fit signal strength measurements $\sigma/\sigma_{SM}=\cos^2\theta=0.87\pm 0.23$. In this plot, there is always a region satisfying $123\, {\rm GeV}\leq m_h \leq 129\, {\rm GeV}$ and the region for singlet mass is $ m_{\sigma} \leq 108 \,{\rm GeV}$. When $\lambda>-0.15$ we have $\delta m_{\sigma}^2\geq 0$. Therefore, for $\lambda<-0.15$, introducing vector fermions entails the undesired effect of making the negative $\delta m_{\sigma}^2$ even more negative.}
\label{fig:Veltman2}
\end{figure}

The mass of the physical field $\sigma$ is also shifted via the quadratically divergent quantum corrections.

\bea\label{eq:delmphi2}
\delta m_{\sigma}^2\!\!\!&=&\!\!\frac{\Lambda^2}{16 \pi^2}\bigg[\!\cos^4\!\theta\left(\!\frac{\lambda}{2}\!\!+3\lambda_S\!\right)\!\!+\!\sin^4\!\theta\left(\!\frac{\lambda}{2}+\!3\lambda_H\!\right)\!\non\\
&+&\!\sin^2\!2\theta\left(\frac{3\lambda_H}{4}\!\!+\!\!\frac{3\lambda_S}{4}\!\!+\!\!\frac{\lambda}{4}\!\right)\!
+\cos^2\!\theta\!\left(\frac{3}{2}\lambda\right)\!\non\\
&+&\sin^2\!\theta\!\left(3\!\lambda_H\!\!+\!\frac{9}{4}g^2\!\!+\!\frac{3}{4}g'^2\!\!\!\!\!-\!\!6g_t^2\!\right)\bigg]\!.
\eea
%
%
In the limit of no mixing $\delta m_{\sigma}^2$ reduces to  $\frac{\Lambda^2}{16\pi^2}(2\lambda+3\lambda_S)$, which well agrees with the results in \cite{Kundu:1994bs}. 
%
%
%
To illustrate the change in the VC, satisfying the experimental bounds, we plot in Figures \ref{fig:Veltman1} and \ref{fig:Veltman2}, $\delta m_h^2/\Lambda^2$ and $\delta m_{\sigma}^2/\Lambda^2$, respectively, as a function of the mixing angle $\theta$ and the quartic coupling $\lambda$. Both plots are produced for $\lambda_H=\lambda_S=0.1$ (keeping $\lambda_S$ small ensures the new scalar to be lighter than the Higgs boson), and the range of other parameters are set based on the following constraints:
\begin{itemize}
\item The observed best-fit signal strength at the LHC $\sigma/\sigma_{SM}=\cos^2\theta=0.87\pm 0.23$ \cite{Chatrchyan:2012ufa} restricts the allowed region for the mixing angle to be either in the interval $0\leq\theta\leq\pi/5$ for the positive roots or $4\pi/5\leq\theta\leq\pi$ for the negative roots of $\cos^2\theta$.
\item The mass of the fundamental scalar discovered at the LHC is found to be at $m_h=125.9\pm 0.4$ GeV \cite{Aad:2012tfa, Chatrchyan:2012ufa}. The figures below are produced by slightly extending this bound to be in the range $123 \,{\rm GeV}\leq m_h \leq 129\, {\rm GeV}$, which restricts the range of quartic coupling to be $-0.9\leq \lambda\leq 0.08$.
\item Constraints from electroweak precision observables (EWPO), are taken into account which favour the mass of the heavier scalar in an extended SM scenario with a real singlet to be below 220 GeV, when the mixing is maximal \cite{Profumo:2007wc}. In Ref. \cite{Barger:2007im} the ranges of the parameters are given as $-3\leq\lambda\leq 3$ and $0\leq \lambda_H\leq 3$, based on the detailed analysis of the branching fractions of the heavier scalar which agrees well with the parameter space chosen in our model. Moreover, in Table 2 of Ref. \cite{Barger:2007im}, the parameter space is defined for 30 fb$^{-1}$ of data from CMS, which are respected in defining the ranges of the parameters used in the following figures.
\item The physical states are ensured by the condition  $\upsilon_{SH}=\upsilon_S/\upsilon_H>0$ which restricts the mixing angle to be either in the interval $0\leq\theta<\pi/4$ or in $\pi/2\leq\theta<3\pi/4$. Note that, as it follows from the Eqs. (\ref{eq:tantheta}) and (\ref{eq:mh2msig2}), $\upsilon_{SH}$ is undefined at $\theta=\pi/4$ and $\,3\pi/4$, so as the masses of the physical scalars. We exclude these points in our analysis.
\item The combination of the constraints by the LHC (first bullet) and the requirement that $\upsilon_{SH}>0$ (fourth bullet), together restrict the range of mixing angle to be $0\leq\theta\leq\pi/5$, which prescribes the parameter space of $\theta$ into a narrow region.

\item The parameter $\lambda_S$, restricts the parameter space by the condition that, the physical masses of the scalars to be real, which is satisfied when $16\lambda_H\lambda_S>\lambda^2$. In our plots, we take $\lambda_H=\lambda_S=0.1$, which gives $\lambda^2<0.16$ or $-0.4<\lambda<0.4$. This constraint is shown in our plots as light-green shaded areas where $m_{\sigma}^2$ becomes negative when $\lambda<-0.4$.
\end{itemize}
In the following figures, the shaded regions are the exclusions, namely, the light-gray shaded region is where the condition $\upsilon_{SH}=\upsilon_S/\upsilon_H>0$ cannot be satisfied, the light-green shaded region is because $m_{\sigma}^2 > 0$ is violated, and light-orange region is ruled out at the LHC via the best-fit signal strength measurements. Within the allowed parameter space, the range for the real singlet mass is $m_{\sigma}\leq 108\, {\rm GeV}$, and there is always a parameter space satisfying $123 \,{\rm GeV}\leq m_h \leq 129\, {\rm GeV}$. In Figure \ref{fig:Veltman1}, the contours represent $\delta m_h^2$ in the units of $\Lambda^2$. When $\lambda_S$ is increased, the contours in Figure \ref{fig:Veltman1} shift to the left. As it is apparent from Figure \ref{fig:Veltman1}, the quantum corrections to the Higgs mass squared, due to the presence of only one additional real singlet scalar has the correct sign, decreasing the negativity. However, the allowed parameter space is very much restricted via the experimental and phenomenological constraints.
It can be inferred from this figure that, the larger the values of $\lambda$, the closer we are to the solution of fine-tuning problem, and the allowed range for the mixing angle is $0\leq\theta\leq \pi/5$. Note that, the dark red, thick, solid contour line, having $\delta m_h^2/\Lambda^2=-0.027$, intersects with the point at $\theta=0$ and $\lambda=0$, and well agrees with the pure SM result on radiative corrections to the Higgs mass. We have the zero contour line, indicated by the thin red solid curve. However, it is excluded both by the LHC best-fit signal strength measurements $\sigma/\sigma_{SM}=\cos^2\theta=0.87\pm 0.23$, and by the condition that $\upsilon_{SH}=\upsilon_{S}/\upsilon_H\geq 0$. The exact cancellation could be achieved via the addition of more scalars \cite{Kundu:1994bs}.

In Figure \ref{fig:Veltman2}, on the other hand, the contours represent $\delta m_{\sigma}^2$  in units of $\Lambda^2$, and the contour lines on the upper and lower halves, shift towards the center if $\lambda_S$ is increased.
The quantum corrections of the mass squared of the new physical state comes with negative sign for the smaller values of $\lambda$, and turns positive if $\lambda>-0.15$.

\begin{figure}[h]
\includegraphics[scale=0.74]{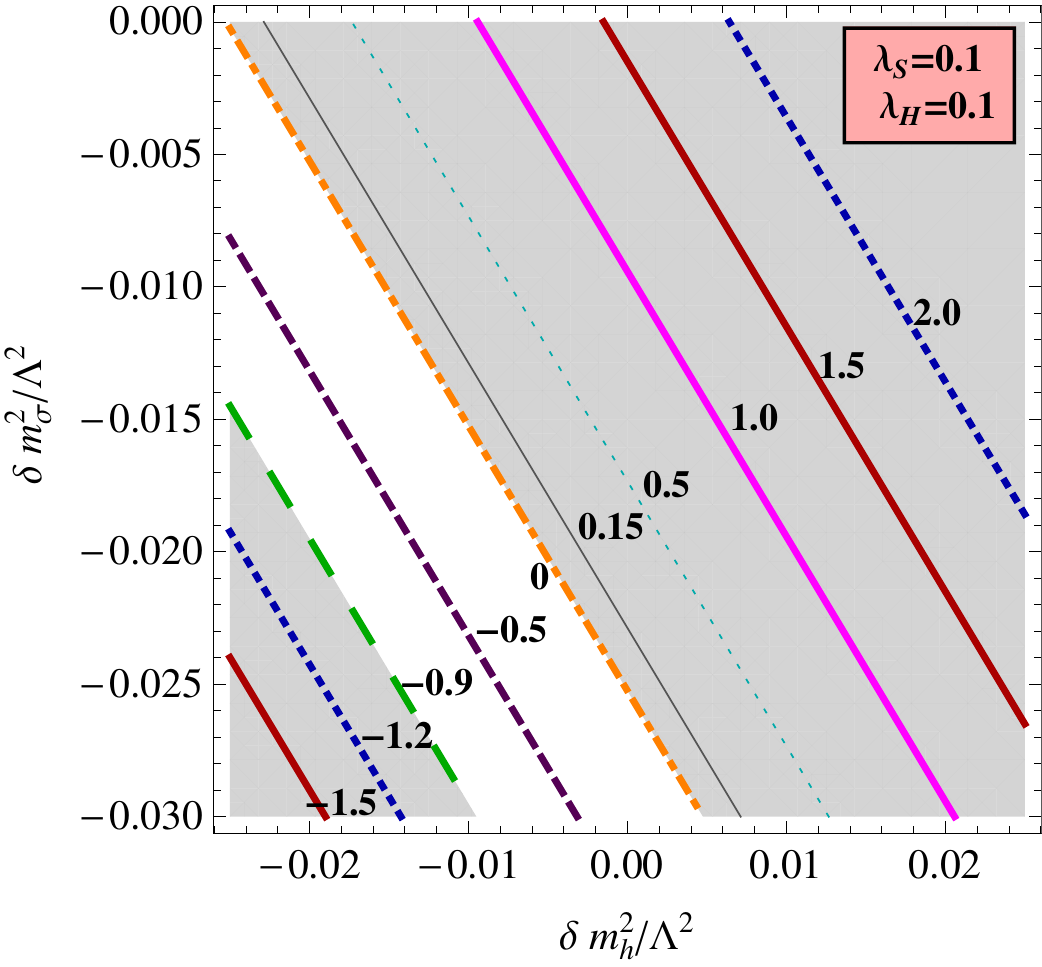}
\caption{The quadratic divergences in the mass squared values of  Higgs and real singlet scalar in unit of $\Lambda^2$ as a function of $\lambda$. The plot is produced for $\lambda_H=\lambda_S=0.1$ and $\theta=4\pi/25$. The ranges for $\delta m_h^2/\Lambda^2$ and $\delta m_{\sigma}^2/\Lambda^2$ are taken to be consistent with Figures \ref{fig:Veltman1} and \ref{fig:Veltman2}. Contours indicate the values of $\lambda$. The light-gray shaded regions are excluded because the interval 123 GeV $<m_h<$ 129 GeV cannot be satisfied in these regions. There is no allowed parameter space satisfying the cancellation of quadratic divergences in both the Higgs and the real singlet masses, simultaneously.}
\label{fig:Veltman3}
\end{figure}

In Figure \ref{fig:Veltman3}, we plot $\delta m_h^2/\Lambda^2$ and $\delta m_{\sigma}^2/\Lambda^2$ as a function of the quartic coupling $\lambda$. We choose $\lambda_H=\lambda_S=0.1$ and $\theta=4\pi/25$ (as a point in the allowed parameter space of both $\delta m_h^2$ in Figure \ref{fig:Veltman1}, and of $\delta m_{\sigma}^2$ in Figure \ref{fig:Veltman2}). Changing the mixing angle does not shift the positions of the contour lines, however, the allowed parameter space for $\lambda$ is different for different $\theta$ values. The larger values of $\theta$ requires $\lambda$ to be negative for satisfying the Higgs mass bound $123\, {\rm GeV}\leq m_h \leq 129\, {\rm GeV}$. The light-gray shaded regions are the exclusions valid for all allowed parameter space of $\theta$ (i.e., $0\leq\theta\leq\pi/5$ and $4\pi/5\leq\theta\leq\pi$). It is clear from this figure that, there is no allowed parameter space to satisfy the cancellation of the quadratic divergences in both Higgs and the new scalar mass squares, simultaneously.

In the literature, the common attempt to get rid of the quadratic divergences in the mass squared of the additional scalar is to introduce either singlet or doublet vector fermions (the chiral fermions do not couple to singlet scalar $S$), which brings a negative contribution to $\delta m_{\sigma}^2$. The corresponding Lagrangian density for vector fermions reads
\bea
\mathcal{L}_{\psi}=\bar{\psi}i \not\!\!{D} \psi -\mu_{\psi}\bar{\psi}\psi-\lambda_{\psi}\bar{\psi}\psi S.
\eea
Note that, $S$ being odd under ${\mathbb{Z}}_2$ symmetry calls for $\psi\rightarrow i\gamma_5\psi$, and the physical mass of the vector fermion is $m_{\psi}=\mu_{\psi}+\lambda_{\psi}\upsilon_S$. Introduction of such fermions in the scenario with mixed scalar states worsens the naturalness problem in the mass squared of the new physical scalar for $\lambda<-0.15$. By the same token, on the grounds that there is mixing between the CP-even scalar of the Higgs doublet and real singlet scalar $S$, when transformed into the physical states, there emerges tree-level coupling of the Higgs boson to vector fermions, which makes the already negative $\delta m_h^2$ even more negative, and therefore, fine tuning problem in the Higgs mass with the additional vector fermions also gets worse.

It is important to note that the results presented here are conclusive only for one loop effects. Higher order corrections might bring a marginal change. We refer reader to \cite{Drozd:2011aa} for higher order corrections.

\section{Conclusion}
Now that the main task of the LHC, to shed light on our understanding of the electroweak symmetry breaking mechanism, has been successfully accomplished with the discovery of the Higgs boson \cite{Aad:2012tfa, Chatrchyan:2012ufa}, the naturalness problem in its mass has become essential to comprehend. To address the issue, we consider the simplest extension of the SM with an additional real singlet scalar, which develops a VEV after electroweak symmetry breaking, and interacts with the SM fields via its direct coupling to the SM Higgs doublet. The mixing between the neutral, CP-even component of the SM Higgs doublet, and the additional field $S$, allows the physical state $\sigma$ to have tree level couplings to the SM fields.

We found that, the minimal modification of the SM with the addition of one real singlet scalar has the right effect on $\delta m_h^2$, decreasing the negativity, though one needs to introduce more scalars to cancel the divergence completely, since the effect is very little in the allowed parameter space. The quadratic divergence on $\delta m_{\sigma}^2$, also comes with an overall negative coefficient for $\lambda<-0.15$. Therefore, the typical venture to cancel the divergences in the singlet scalar mass by introducing vector fermions might not be a solution for the additional scalar sector when $\lambda<-0.15$, and it seems not a viable scenario from the Higgs boson point of view in the case of mixing between the scalars, as it enhances $\delta m_h^2$, too. To conclude, we can state that it is difficult, if not
impossible, to naturalize the SM Higgs boson by coupling it to singlet scalars, and that the additional vector fermions in this scenario might entail the undesired effect of enhancement in both $\delta m_h^2$ and $\delta m_{\sigma}^2$.
\section{Appendix}
Below we show the vertex factors used in this work.
\begin{flalign*}
&\lambda_{\rm hhhh}=-6i\left[\lambda_H\cos^4\theta+\lambda_S\sin^4\theta+\frac{\lambda}{4}\sin^22\theta\right]\!,&\\
&\lambda_{\sigma\sigma\sigma\sigma}=-6i\left[\lambda_H\sin^4\theta+\lambda_S\cos^4\theta+\frac{\lambda}{4}\sin^22\theta\right]\!,&\\
&\lambda_{\textrm{hh}\sigma\sigma}=-\frac{i}{2}\left[\!\left(\!3\lambda_H\!+\!3\lambda_S\!-\!2\lambda\!\right)\!\sin^2\!2\theta
\!+2\!\lambda(\sin^4\!\!\theta\!+\!\cos^4\!\!\theta)\!\right]\!,&\\
&\lambda_{\textrm{h}(\sigma)W_{\mu}^+W_{\nu}^-}=\frac{i}{2}g^2\upsilon_Hg^{\mu\nu}\cos\theta(\sin\theta),&\\
\end{flalign*}
\begin{flalign*}
&\lambda_{\textrm{hh}(\sigma\sigma) W_{\mu}^+W_{\nu}^-}=\frac{i}{2}g^2 g^{\mu\nu}\cos^2\theta(\sin^2\theta),&\\
&\lambda_{\textrm{h}(\sigma)Z_{\mu}Z_{\nu}}=\frac{i}{2}g_Z^2\upsilon_Hg^{\mu\nu}\cos\theta(\sin\theta),&\\
&\lambda_{\textrm{hh}(\sigma\sigma) Z_{\mu}Z_{\nu}}=\frac{i}{2}g_Z^2 g^{\mu\nu}\cos^2\theta(\sin^2\theta),&\\
&\lambda_{\rm hh(\sigma\sigma)H_{2}H_{2}}\!=\!-i\!\left[2\lambda_H\cos^2\!\theta(\sin^2\!\theta)\!+\!
\lambda\sin^2\!\theta(\cos^2\!\theta)\right]\!,&\\
&\lambda_{\rm hh(\sigma\sigma)H_{3}H_{3}}\!=\!-i\!\left[2\lambda_H\cos^2\!\theta(\sin^2\!\theta)\!+\!
\lambda\sin^2\!\theta(\cos^2\!\theta)\right]\!,&\\
&\lambda_{\rm hh(\sigma\sigma)H_{4}H_{4}}\!=\!-i\!\left[2\lambda_H\cos^2\!\theta(\sin^2\!\theta)\!+\!
\lambda\sin^2\!\theta(\cos^2\!\theta)\right]\!,&\\
&\lambda_{h(\sigma)t\bar{t}}=\frac{-i}{\sqrt{2}}\frac{m_t}{\upsilon_H}\delta_{ab}\delta_{\alpha\beta}\cos\theta(\sin\theta),&
\end{flalign*}
%
where  $g_Z^2=g^2+g'^2$, and $g$ and $g'$ are $SU(2)_L$ and $U(1)_Y$ gauge couplings of the SM, respectively. The subscripts $a$, $b$ appearing in the last vertex factor represent the color indices, and the subscripts $\alpha$, $\beta$ are the spinor indices.\\

\noindent\textbf{Acknowledgments}\\

We thank D. A. Demir for fruitful discussions. We are grateful to the very conscientious referee for his/her constructive comments and suggestions.

This work has been supported by T\"{U}B\.{I}TAK, The Scientific and Technical Research Council of Turkey, through the grant 2232, Project No: 113C002.
%

\end{document}